\documentstyle[prb,aps,epsfig,multicol,amstex]{revtex}

\begin{document}
\preprint{MA/UC3M/14/1998}
\draft
\title{Fundamental measure theory for mixtures of parallel hard cubes. \\
II. Phase behaviour of the one-component fluid and of the binary mixture}
\author{Yuri Mart\'\i nez-Rat\'on\cite{emailyuri} and
	Jos\'e A.\ Cuesta\cite{emailjose}}
\address{Grupo Interdisciplinar de Sistemas Complicados
(GISC), Departamento de Matem\'aticas, Escuela Polit\'ecnica
Superior, Universidad Carlos III de Madrid, Avda.\ de la Universidad, 30,
28911 -- Legan\'es, Madrid, Spain}

\maketitle

\begin{abstract}
A previously developed fundamental measure functional [\jcp {\bf 107},
6379 (1997)] is used to study the phase behaviour of a
system of parallel hard cubes. The single-component fluid exhibits
a continuous transition to a solid with an anomalously large density
of vacancies. The binary mixture has a demixing transition
for edge-length ratios below 0.1. Freezing in this mixture
reveals that at least the phase rich in large cubes always lies in
the region where the uniform fluid is unstable, hence suggesting a
fluid-solid phase separation. A method is developed to study very
asymmetric binary mixtures by taking the limit of zero
size ratio (scaling density and fugacity of the solvent as
appropriate) in the semi-grand ensemble where the chemical potential
of the solvent is fixed. With this procedure the mixture is exactly
mapped onto a one-component fluid of parallel adhesive hard cubes.
At any density and solvent fugacity the large cubes are shown to
collapse into a close-packed solid. Nevertheless the phase diagram
contains a large metastability region with fluid and solid phases.
Upon introduction of a slight polydispersity in the large cubes the
system shows the typical phase diagram of a fluid
with an isostructural solid-solid transition (with the exception
of a continuous freezing). Consequences about the phase behaviour
of binary mixtures of hard core particles are then drawn.
\end{abstract}

% 61.20.Gy : Theory and models of liquids
% 64.75.+g : Solubility, segregation, and mixing; phase separation
% 82.70.Dd : Colloids

\begin{multicols}{2}
\narrowtext

\section{Introduction}
\label{introduction}

This paper is the sequel of a previous one\cite{cuestaI}
(henceforth referred to as I) in which the so-called
{\em fundamental measure theory} (FMT) was applied to build a density
functional for the multicomponent system of parallel hard cubes (PHC).
In I we explained all fundamentals of the theory and gave a full
account of the technical details involved in the derivation of the
functional. We also discussed the pros and cons of the theory, as
compared with more standard density functional theories (DFTs), and
suggested some possible extensions.

In this paper we apply the formalism developed in I to study the 
phase behaviour of the PHC fluid, with special emphasis in its relevance
for the understanding of the phase behaviour of mixtures. The PHC
fluid is a rather academic model and it possesses a bunch of `peculiarities'
which are rather odd for a fluid model, e.g.\ the uniform fluid is
anisotropic
at small scales (the cubes are kept parallel to each other), freezing
occurs at very low packing fractions (around 0.3--0.4) and it is a
continuous, instead of first order, transition (a consequence of the
lack of isotropy of the fluid phase), and the depletion in the binary
mixture is very strong compared to hard spheres (HS).
But in spite of these peculiarities---which certainly make of this
model a caricature of a fluid, the
physics one can learn from its phase behaviour can be easily extended to
more reasonable fluids, and it has the important added value of being a
much simpler model to carry out analytical calculations (even more: FMT
seems to be somehow optimal for this model \cite{cuestaI,cuesta1}).

There have been a few previous studies in the literature about the fluid
of PHC \cite{zwanzig,hoover1,hoover2,vanswol,jagla,kirkpatrick} scattered in
the last forty years, but the relevance of this model has only
recently become apparent when it has been proposed as a model of a
fluid able to demix by a purely entropic mechanism. \cite{dijkstra,cuesta2}

Entropic demixing has been a long standing question which only 
recently begins to be understood. It is well known that different
attractions between two types of particles in a fluid can produce
segregation into two phases, each rich in one type of particles.
\cite{rowlinson} The question remains whether hard particles,
for which only an entropic balance can drive a phase transition, ever
demix and how. It is clear that nonadditive mixtures (mixtures in
which particles of different type interact as if they had a larger
volume) do demix, \cite{melnyk,adams,ehrenberg,carmesin} but they have the
segregation mechanism introduced at the interaction potential. So
the nontrivial question concerns additive hard-particle mixtures.
The question is tricky because the simplest model
of this type---HS---was solved in Percus-Yevick (PY)
approximation \cite{lebowitz1} and shown never to demix. \cite{lebowitz2}
It had to wait almost thirty years until the PY result was questioned.
By solving numerically the Ornstein-Zernike equation with closure relations
more accurate than the PY one a spinodal instability was shown to
occur in a binary mixture of HS
\cite{biben2,lekkerkerker,rosenfeld1} for a diameter ratio below
0.1--0.25 (depending of the authors).

It was then believed that a sufficiently asymmetric HS
binary mixture undergoes fluid-fluid demixing in a certain region of
the phase diagram. But successive experiments performed in suspensions
of polystyrene or silica spheres (which to a large accuracy can be
considered HS) showed cumulative evidence that demixing
is coupled to freezing, and so one of the phases shows up as a
crystal---sometimes a glass---of large particles.
\cite{duijneveldt,kaplan,steiner,imhof} Recent theoretical calculations
confirm this scheme. \cite{poon,caccamo}

Direct computer simulations have run into troubles when dealing
with this problem due to
the extremely low probability of moving a larger particle in a sea
of small ones without overlapping any of them. Simulations are still
possible for not too dissimilar diameters, but they do not
show demixing. \cite{fries,jackson} With the help of specially designed
cluster moves the first evidence of demixing has recently been found in
simulations for diameter ratios smaller than 0.05. \cite{buhot}
Unfortunately these cluster moves do not help above the percolation
threshold, and this sets a relatively low upper bound for the total
packing fraction of the fluid. Besides, this method does not allow to
identify the coexisting phases, so it does not distinguish
fluid-fluid or fluid-solid demixing.

%There is, however, indirect evidence
%for the latter: extrapolating from the simulations of 
%binary mixtures of hard spherocylinders to the limit of zero cylinder
%length it has been shown \cite{dijkstra2} fluid-solid demixing
%to occur in the HS binary mixture for diameter ratios
%below 0.1.

In order to elucidate the nature of demixing for very asymmetric
mixtures the attention has shifted to understand the depletion
interaction.  An effective pair potential between the large spheres can
be obtained by different procedures. \cite{mao,biben3,noe,dijkstra3}
Its shape reveals a very deep and narrow (one small-sphere diameter)
well followed by a couple of small oscillations extending two or
three small-sphere diameters. The depth and the amplitude of the 
oscillations depend on the small spheres packing fraction. In view
of the phase behaviour of spheres with narrow and deep attractive
potentials \cite{bolhuis,tejero} fluid-fluid demixing is ruled out
for sufficient asymmetry, and instead fluid-solid demixing or even
expanded-dense solid demixing (i.e.\ demixing once the large component
has crystallised) should appear. Simulations of a system of HS 
supplemented by the depletion potential \cite{biben3,noe,dijkstra3}
confirm this conjecture if the diameter ratio is below 0.1. 
Expanded-dense solid demixing is indeed shown in some of this 
simulations for ratios 0.1 \cite{noe} or 0.05. \cite{dijkstra3}
This phase behaviour has been recently corroborated by {\em direct}
simulations of the binary mixture, \cite{dijkstra4} which have
been shown to be possible in the relevant region of the phase
diagram thanks to the small amount of small spheres present
in those statepoints.

In the limit of zero size of the small spheres at constant packing
fraction the depletion potential becomes Baxter's adhesive potential.
\cite{biben1,heno} By applying a similar limit to the FMT functional
of a binary mixture of PHC \cite{cuestaI,cuesta1} we have recently
obtained a functional for parallel adhesive hard cubes (PAHC).
\cite{yuri} By avoiding the singularities of this potential (we will
treat this point in detail later on) we show that the phase behaviour
of this fluid is consistent with the
above described picture for the mixture of HS, with the difference
that expanded-dense solid demixing is the most common scenario 
because PAHC freezing occurs at rather low packing fractions.

\section{The one-component fluid}
\label{one-component}

There are relatively few results in the literature concerning the 
fluid of PHC. As concerns the uniform fluid the virial coefficients,
both in 2D and
3D, have been exactly obtained through diagrammatic expansions up
to the seventh. \cite{zwanzig,hoover1} They have also
been obtained for different approximate integral equation theories.
\cite{hoover2} The main consequence one draws from this information
is that the virial expansion is very poorly convergent. For the 3D
fluid, the seventh
order virial expansion exhibits a maximum in the pressure at $\eta
\approx 0.6$, and then goes down very quickly to reach negative
values. \cite{hoover2} Its behaviour at moderate densities strongly
deviates from that of a sixth order virial expansion---in contrast
with what happens for HS. This is the fingerprint of a
nearby divergence. In fact, the simulations of the fluid of PHC (3D)
\cite{vanswol,jagla} show a continuous freezing into a simple cubic
lattice at \cite{jagla} $\eta=0.48\pm 0.02$
(in contrast with the first order
nature of the HS freezing). This result has been proven to be
exact in infinite dimensions. \cite{kirkpatrick} The reason for
a continuous freezing is the lack of rotational symmetry of this
system even in the disordered phase \cite{kirkpatrick} (thus 
freezing does not brakes this symmetry). Allowing the cubes to
rotate restores the first order nature of the freezing transition.
\cite{jagla}

%there is a simulation on
%the fluid of PHC (3D) \cite{vanswol} where it is shown that freezing
%occurs around $\eta\approx 0.4$--0.5; unfortunately (see Fig.\
%\ref{eqofstate}) there are too few state points in this region as
%to ascertain the order of the transition. It has been reported 
%that PHC undergo a continuous freezing into a simple cubic
%lattice---again in contrast with the first order nature of
%HS freezing---in infinite dimensions. \cite{kirkpatrick}
%The physical reason why this is so is that the PHC fluid is anisotropic
%at small scales and thus rotational symmetry is broken already at the
%disordered phase. \cite{kirkpatrick} It is therefore reasonable to
%expect a similar behaviour at any dimension, and in fact the simulations
%are compatible with a continuous freezing.

\begin{figure}
\epsfig{file=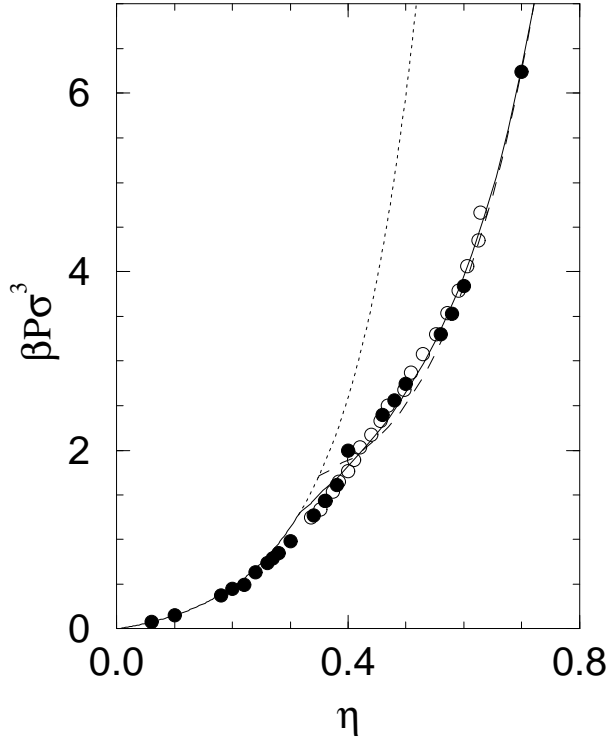, width=3.2in}
\caption[]{Equation of state of the PHC fluid. Solid lines correspond
to the pressure of the stable phase (fluid or solid) at the given
packing fraction, $\eta$. Dotted line is the unstable fluid branch
beyond freezing. Dashed line is the solid branch computed without
accounting for vacancies in the lattice (see text). Full circles are
the simulations of Ref.\ \onlinecite{vanswol} and empty circles
those of Ref.\ \onlinecite{jagla} (actually we have averaged
the two data sets reported for the largest system size).}
\label{eqofstate}
\end{figure}

FMT's equation of state of the uniform PHC fluid is
simply that of the scaled particle theory (SPT), i.e., from Eqs.\ (60)
and (61) of I,
\begin{eqnarray}
\beta P/\rho &=& \frac{1}{(1-\eta)^2}\, , \quad (D=2), \label{eos2D} \\
\beta P/\rho &=& \frac{1+\eta}{(1-\eta)^3}\, , \quad (D=3), \label{eos3D}
\end{eqnarray}
with $\beta$ the inverse absolute temperature in Boltzmann constant
units, $P$ the pressure, $\rho$ the number density, and $\eta$ the
packing fraction, $\eta=\rho\sigma^D$, $\sigma$ being the cube
edge-length.  In Fig.\ \ref{eqofstate} the SPT equation of state
(\ref{eos3D})
is compared with the simulation data.

On the other hand freezing can
be studied as usual in DFT by parametrising the local density with a
sum of gaussians centered at the lattice points. \cite{evans} Since
the lattice is simple cubic, this can be easily achieved by setting
\begin{mathletters}
\begin{eqnarray}
\rho({\bf r}) &=& \prod_{\nu=x,y,z}
\sum_{n_{\nu}=-\infty}^{\infty}g(\nu-n_{\nu}d) \, ,
\label{gxgygz}  \\
g(u) &\equiv& \left(\frac{\alpha}{\pi}\right)^{1/2}e^{-\alpha u^2}
\, , \label{gaussians}
\end{eqnarray}
\label{densprof}
\end{mathletters}

\noindent
with $d=\eta^{-1/3}\sigma$ the lattice spacing, and $\alpha$ a variational
parameter determining the localisation of particles at the lattice sites.

In I we expressed the free-energy functional of a multicomponent system
with density profiles $\rho_i({\bf r})$ ($i$ labeling the species) as 
\begin{eqnarray}
F[\{\rho_i({\bf r})\}] &=& F^{\rm id}[\{\rho_i({\bf r})\}]+
F^{\rm ex}[\{\rho_i({\bf r})\}] \, ,  \label{Ftotal}  \\
\beta F^{\rm id}[\{\rho_i({\bf r})\}] &=& \sum_i\int d{\bf r}\,
\rho_i({\bf r})\left[\ln{\cal V}_i\rho_i({\bf r})-1\right]  \, ,
\label{Fideal}  \\
\beta F^{\rm ex}[\{\rho_i({\bf r})\}] &=& \int d{\bf r}\,
\Phi(\{n_{\alpha}({\bf r})\})  \, ,  \label{Fexcess}
\end{eqnarray}
with ${\cal V}_i$ the thermal volume of species $i$, and
$n_{\alpha}=\sum_i\rho_i*\omega_i^{(\alpha)}$ a set of four
weighted densities (stars denotes convolution) defined by the weights
\begin{mathletters}
\begin{eqnarray}
\omega^{(0)}_i &\equiv& \delta_i^x\delta_i^y\delta_i^z \, ,
\label{omega0} \\
{\bf w}^{(1)}_i &\equiv&
     \bigl(\theta_i^x\delta_i^y\delta_i^z,\delta_i^x\theta_i^y\delta_i^z,
     \delta_i^x\delta_i^y\theta_i^z\bigr) \, ,
\label{omega1} \\
{\bf w}^{(2)}_i &\equiv&
     \bigl(\delta_i^x\theta_i^y\theta_i^z,\theta_i^x\delta_i^y\theta_i^z,
     \theta_i^x\theta_i^y\delta_i^z\bigr) \, ,
\label{omega2} \\
\omega^{(3)}_i &\equiv& \theta_i^x\theta_i^y\theta_i^z \, ,
\label{omega3}
\end{eqnarray}
\label{omegas}
\end{mathletters}

\noindent
with $\theta_i^u\equiv\Theta(\sigma_i/2-|u|)$, and
$\delta_i^u\equiv(1/2)\delta(\sigma_i/2-|u|)$.
As in I we will also need the two scalar
densities $n_i=n_{i,x}+n_{i,y}+n_{i,z}$, $n_{i,\nu}$, $\nu=x,y,z$,
being the vector components of ${\bf n}_i$, $i=1,2$. The function
$\Phi({\bf r})$ is given in terms of the $n_{\alpha}$'s as (see I)
\begin{equation}
\Phi({\bf r})=-n_0\ln(1-n_3)+
\frac{{\bf n}_1\cdot{\bf n}_2}{1-n_3}+\frac{n_{2,x}n_{2,y}n_{2,z}}
{(1-n_3)^2}
\label{excess}
\end{equation}
with $n_{2,\nu}$, $\nu=x,y,z$, the components of ${\bf n}_2$.

For convenience let us introduce the functions 
\begin{eqnarray}
e(u) &\equiv& \frac{1}{2}\mbox{erf}\,\left(\sqrt{\alpha}u\right) \, ,
\label{functione}  \\
p(u) &\equiv & \sum_{n=-\infty}^{\infty}[e(u-nd+\sigma/2)-e(u-nd-\sigma/2)]
\, , \label{functionp}  \\
q(u) &\equiv & \sum_{n=-\infty}^{\infty}[g(u-nd+\sigma/2)+g(u-nd-\sigma/2)]
\label{functionq}
\end{eqnarray}
[notice that $g(u)=e'(u)$]. The latter two are periodic with period
$d$. In terms of these functions
\begin{mathletters}
\begin{eqnarray}
n_3 &=& p(x)p(y)p(z) \, ,  \label{n3gauss} \\
{\bf n}_2 &=& \frac{1}{2}\bigl(q(x)p(y)p(z),p(x)q(y)p(z),
p(x)p(y)q(z)\bigr)  \, ,   \label{n2gauss} \\
{\bf n}_1 &=& \frac{1}{4}\bigl(p(x)q(y)q(z),q(x)p(y)q(z),
q(x)q(y)p(z)\bigr)  \, ,   \label{n1gauss} \\
n_0 &=& \frac{1}{8}q(x)q(y)q(z) \, ,  \label{n0gauss}
\end{eqnarray}
\label{ngauss}
\end{mathletters}

\noindent
and accordingly $\Phi({\bf r})=n_0({\bf r})\psi({\bf r})$, where
\begin{equation}
\psi=-\ln(1-n_3)+\frac{n_3}{1-n_3}+\frac{n_3^2}{(1-n_3)^2} \, .
\label{psi}
\end{equation}
This simplification occurs because of the factorisation of the density
(\ref{gxgygz})---and hence of (\ref{ngauss}).

The free energy {\em per particle}, $\Psi\equiv\beta F/N$,
of the solid phase can be obtained as the integral over a unit cell
of the lattice of the free energy density. Hence the ideal part
turns out to be
\begin{eqnarray}
\Psi^{\rm id} &=& \int_{-d/2}^{d/2}dx
\int_{-d/2}^{d/2}dy\int_{-d/2}^{d/2}dz\,\rho({\bf r})[\ln{\cal V}
\rho({\bf r})-1]  \nonumber  \\
&=& \ln({\cal V}/\sigma^3)-1   \nonumber  \\
&&+3\int_{-\infty}^{\infty}dx\,g(x)\ln\left(
\sum_{n=-\infty}^{\infty}\sigma g(x-nd)\right) \, .
\label{FNid}
\end{eqnarray}
The integrand is written in a suitable way to use Gauss-Hermite
numerical quadratures. \cite{press}
As for the excess contribution, similar manipulations lead to
\begin{eqnarray}
\Psi^{\rm ex} &=& \frac{1}{8}\int_{-d/2}^{d/2}dx\,q(x)
\int_{-d/2}^{d/2}dy\,q(y)\int_{-d/2}^{d/2}dz\,q(z)\psi(x,y,z)
\nonumber  \\
&=& \frac{1}{8}\int_{-\infty}^{\infty}dx\,g(x)
\int_{-\infty}^{\infty}dy\,g(y)
\int_{-\infty}^{\infty}dz\,g(z)  \nonumber \\
&&\times\sum_{\{\pm\}}\psi(x\pm\sigma/2,y\pm\sigma/2,z\pm\sigma/2) \, ,
\label{FNex}
\end{eqnarray}
where the last summation runs over all combinations of signs.
Again the latter expression is suitable for Gauss-Hermite
integration, which is crucial this time because (\ref{FNex})
involves a three-dimensional integration.

We can now minimise with respect to $\alpha$ to determine the
equilibrium profile. This yields a continuous freezing at $\eta=0.348$.
As the transition is continuous we can make a more accurate
determination of the transition
density via a standard bifurcation analysis. This is equivalent to
finding the density at which the structure factor diverges for some
wavevector ${\bf k}_c$. The structure factor is expressed in terms
of the DCF as
\begin{equation}
S({\bf k})=\frac{1}{1-\rho\hat c({\bf k})} \, ,
\label{structure}
\end{equation}
and the Fourier transform of the DCF, $\hat c({\bf k})$, is obtained
through Eq.\ (56) of I for a one-component fluid. In order to simplify
the final expression we exploit the symmetry of the crystal by choosing
${\bf k}_c=(k_c,0,0)$ (the result would be the same if we chose $k_c$
along the Y or Z axes). Thus the condition to determine the critical
point is 
\begin{eqnarray}
1&+&2\eta_c\frac{4-3\eta_c+\eta_c^2}{(1-\eta_c)^3}j_0(k\sigma) 
\nonumber \\
&+&\eta_c^2\frac{9-4\eta_c+\eta_c^2}{(1-\eta_c)^4}j_0(k\sigma/2)^2
\geq 0 \, ,
\label{inequal}
\end{eqnarray}
the equality holding only for $k=k_c$; $j_0(x)\equiv\sin x/x$ is the
zeroth order spherical Bessel function. The solution to this equation
is $\eta_c=0.3143\dots$ and $k_c\sigma=4.8276\dots$. 

It is noticeable the discrepancy between the value of $\eta_c$ obtained
from the divergence of the structure factor and that obtained using the
profile (\ref{densprof}). The reason for this discrepancy can be inferred
if we obtain from $k_c$ the lattice spacing,
$d_c/\sigma=2\pi/k_c=1.3015\dots$;
thus $\eta_c(d_c/\sigma)^3=0.6929\dots$,
what means that the resulting crystal has a large fraction of vacancies
(around 31\% of the lattice sites!). This is a strong effect that can
be accounted for by simply multiplying the r.h.s.\ of (\ref{gxgygz})
by an average occupancy ratio $\vartheta$ and minimising with respect
to this new variational parameter. In the calculation process this
simply amounts to (i) add a term $\ln\vartheta$ to the ideal free
energy (\ref{FNid}), and (ii) replace $n_3$ by $\vartheta n_3$ in the
definition of $\psi$ [Eq.\ (\ref{psi})]. Notice that $d$ is now given
by $d=\sigma(\vartheta/\eta)^{1/3}$. As a result we obtain the correct
value of $\eta_c$ and an occupancy ratio of $\vartheta\approx 0.694$,
consistent with the value obtained above. \cite{nota1}

The solid equation of state---with and without vacancies---is
plotted in Fig.\ \ref{eqofstate} and compared with the simulations.
It is obtained as $\beta P\sigma^3=\eta^2\partial\Psi/\partial\eta$.
We can see that the overall agreement is good, although the freezing
point is shifted down with respect to the simulations because the 
SPT equation of state overestimates the pressure of the fluid phase.

\section{Stability of the binary fluid mixture}
\label{stability}

Two are the requirements for a fluid mixture to be stable: \cite{callen}
(i) the positiveness of the specific heat at constant volume ($c_V$),
and (ii) the positive definiteness of the matrix
\begin{equation}
{\sf M}_{ij}\equiv\beta\frac{\partial^2f}{\partial\rho_i\partial\rho_j}
\, ,  \label{stab-1}
\end{equation}
where $f\equiv F/V$, $F$ being the Helmholtz free energy, and $\rho_i$
the number density of species $i$.
Condition (i) is trivially fulfilled, because any hard core model is
athermal, what means that the dependence of the free energy on
temperature is that of an ideal gas; hence the positiveness of $c_V$.
Condition (ii) is a consequence
of the equilibrium state being a minimum of the free energy. 

According to (\ref{Fexcess}) the matrix {\sf M} will be given by
\[
{\sf M}_{ij}=\beta\frac{\partial\mu_j}{\partial\rho_i}
 =\frac{1}{\rho_i}\delta_{ij}+\sum_{\alpha\gamma}
  \hat\omega^{(\gamma)}_i({\bf 0})\hat\omega^{(\alpha)}_j({\bf 0})
  \frac{\partial^2\Phi}{\partial n_{\gamma}\partial n_{\alpha}} \, ,
\]
and relating the second term of the r.h.s.\ of this equation to the
DCF of the fluid [Eq.\ (7) of I] it can simply be written as
\begin{equation}
{\sf M}_{ij}=\frac{1}{\rho_i}\delta_{ij}-\hat c_{ij}({\bf 0}) \, .
\label{matrixM}
\end{equation}

Now, for a binary mixture, {\sf M} is a $2\times 2$ matrix with all
its elements positive; thus the mixture will be stable provided
$|{\sf M}|>0$. The solution to the equation $|{\sf M}|=0$, if it
exist at all, will represent a spinodal curve. Such a condition can
be understood in terms of the structure factor matrix of the mixture,
given by
\begin{equation}
\rho{\sf S}({\bf k})\equiv[{\sf P}^{-1}-\hat{\sf C}({\bf k})]^{-1} \, ,
\label{struct}
\end{equation}
where $\hat{\sf C}_{ij}\equiv\hat c_{ij}$ and ${\sf P}_{ij}\equiv
\rho_i\delta_{ij}$. Then, after (\ref{matrixM}), $|{\sf M}|=0$ is the
condition for the structure factor to diverge at zero wavevector (the
uniform fluid).

In order to work out the expression of $|{\sf M}|$ let us introduce the
following notation: $\eta_i\equiv\sigma_i^D\rho_i$, the packing fraction
of species $i$; $\eta\equiv\eta_1+\eta_2=\xi_D$, the total packing
fraction of the fluid; $r\equiv\sigma_1/\sigma_2$, the large-to-small
edge ratio ($r\geq 1$); and $x\equiv\eta_1/\eta$, the relative packing
fraction of the large component. With these definitions as well as the
short-hand ${\cal M}\equiv\rho_1\rho_2|{\sf M}|$ we obtain,
after a tedious but straightforward calculation,
\begin{equation}
{\cal M}=\frac{1+\eta}{(1-\eta)^3} \, ,
\label{det-2D}
\end{equation}
for $D=2$, and \cite{cuesta2}
\begin{equation}
{\cal M}=\frac{\eta^2}{(1-\eta)^4}\left[1+\frac{4}{\eta}
+\frac{1}{\eta^2}-\frac{3(r-1)^2}{r}x(1-x)\right] \, ,
\label{det-3D}
\end{equation}
for $D=3$.

\begin{figure}
\epsfig{file=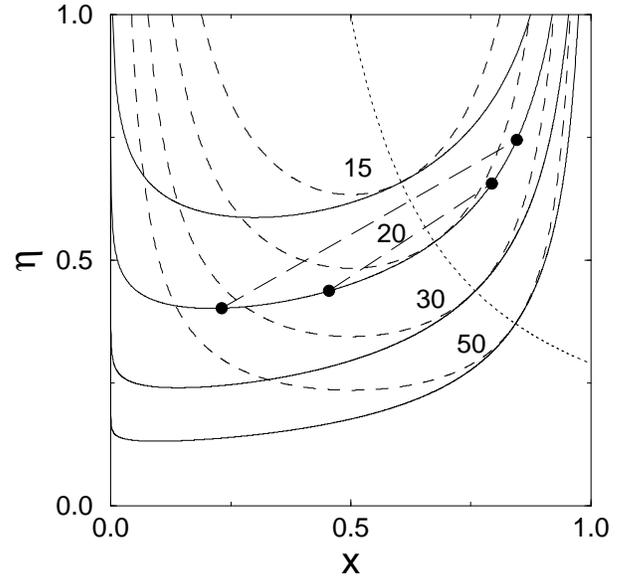, width=3.2in}
\caption[]{Phase diagram of the demixing transition of parallel hard cubes;
$\eta$ is the fraction of volume occupied by all cubes, whereas
$x\equiv\eta_1/\eta$ is the fraction of volume occupied by the large cubes.
Short-dashed lines represent the spinodals for different values of the
edge-to-edge ratio; solid lines are the corresponding coexistence lines
(the actual transition lines); the dotted line is the line of critical
points of the demixing transition for all values of $r>r_c$;
finally, the long-dashed segments joining
black dots are two examples of coexisting states.}
\label{demix}
\end{figure}

From (\ref{det-2D}) it follows that the 2D mixture is stable
whichever the values of $\eta$, $r$ and $x$. Accordingly, parallel
hard squares never demix into two {\em fluids} with different
composition. Equation (\ref{det-3D}), however, tells us
that the mixture of PHC will be stable provided the expression in
square brackets is positive, i.e.\
\begin{equation}
1+\frac{4}{\eta}+\frac{1}{\eta^2}>\frac{3(r-1)^2}{r}x(1-x) \, .
\label{condition}
\end{equation}
Since the minimum of the function $1+4/\eta+1/\eta^2$, for
$0\leq\eta\leq 1$, is 6 (reached when $\eta=1$) and the maximum of
$x(1-x)$, for $0\leq x\leq 1$, is $1/4$ (reached when $x=1/2$),
(\ref{condition}) will hold for {\em any} $\eta$ and $x$ whenever
\[ 6>\frac{3(r-1)^2}{r}\frac{1}{4} \quad \Longleftrightarrow \quad
r^2-10r+1<0 \, , \]
which means for any $1/r_c<r<r_c$, where
\begin{equation}
r_c=5+\sqrt{24}\approx 9.98
\label{rc}
\end{equation}
(or, equivalently, as we have defined $r\geq 1$, for any $1\leq r<r_c$).
For $r\geq r_c$ there will exist values of $\eta$ and $x$ for which
(\ref{condition}) does not hold, and thus the mixture demixes. From
(\ref{condition}) it is very simple to find that those values correspond
to the region above the curve
\begin{equation}
\frac{1}{\eta}=\sqrt{3}\sqrt{1+\frac{(r-1)^2}{r}x(1-x)}-2 \, ,
\label{spinodal}
\end{equation}
which therefore defines the spinodal. Figure \ref{demix} shows this curve
for a few values of $r$. It is interesting to notice the symmetry of
the spinodal with respect to $x=1/2$. This means that for a given packing
fraction, $\eta$, the stability of the mixture depends on the fraction
of occupied volume of any of the particles, regardless their type,
large or small.

The existence of a spinodal instability of this type means that
if the system is kept at constant pressure, there is a certain region
in the density-composition phase diagram in which the fluid is stable
in two coexisting phases, one rich in small particles and the other one
rich in large particles. In order to determine
the values of $\eta$ and $x$ of the two coexisting phases we must solve,
at a given pressure, the equilibrium equations. If we denote $\eta_s$,
$x_s$ (respectively $\eta_{\,l}$, $x_{\,l}$) the values of the small-particle
(respectively large-particle) rich phase, those equations can be written
%\begin{mathletters}
\begin{eqnarray*}
P(\eta_s,x_s) & = & %p \, , \\ %\label{equil-P1}
P(\eta_{\,l},x_{\,l}) = p \, , \\ %\label{equil-P2}
\mu_1(\eta_s,x_s) & = & \mu_1(\eta_{\,l},x_{\,l}) \, , \\ %\label{equil-mu1}
\mu_2(\eta_s,x_s) & = & \mu_2(\eta_{\,l},x_{\,l}) \, , %\label{equil-mu2}
\end{eqnarray*}
%\label{equil}
%\end{mathletters}
$p$ being the externally fixed pressure. This equations express equality
of the pressure and chemical potentials of both kinds of particles in
each of the two coexisting phases. For the present binary mixture,
$\mu_1$, $\mu_2$ and $P$ are given by
\begin{eqnarray}
\beta P\sigma_1^3 &=& y(r^3-(r^3-1)x)+2y^3(r-(r-1)x)^3  \nonumber \\
    & & +3y^2(r^2-(r^2-1)x)(r-(r-1)x) \, ,   \label{bin-press} \\
\beta\Delta\mu_1 &=& \ln(xy)+3y(r^2+r-(r^2+r-2)x)   \nonumber \\
    & & +3y^2(r-(r-1)x)^2+\beta P\sigma_{1}^3 \, ,    \label{bin-mu1} \\
\beta\Delta\mu_2 &=& \ln[(1-x)y]+3y(2-(2-r^{-1}-r^{-2})x)  \nonumber \\
    & & +3y^2(1-(1-r^{-1})x)^2+\beta P\sigma_1^3r^{-3} \, ,  \label{bin-mu2}
\end{eqnarray}
where $y=\eta/(1-\eta)$, and $\beta\Delta\mu_i=\beta\mu_i-\ln(\Lambda_i/
\sigma_i)^3$.

To determine the critical point of this transition we use the fact that
this point is the only one for which the spinodal and the coexistence
line coincide, and it corresponds to the minimum value of the pressure
on the spinodal. Hence Eqs.\ (\ref{spinodal}) and (\ref{bin-press})
allow us to determine this point for every value of the edge-ratio $r$.

The phase diagram (Fig.\ \ref{demix}) shows a few features that are
worth noticing. First of all, it is interesting to see that the
critical line collides, when $r\to\infty$, with the $x=1$ edge of the
phase diagram at a nonzero packing fraction. What this suggests is that
in this limit the packing fraction of the small component
goes to zero but it still remains a residual depletion between the large
cubes. This depletion forces the one-component effective fluid of large
cubes to collapse beyond a certain packing fraction. We will explore
this matter in full detail in Sec.\ \ref{adhesive}.

Another interesting feature of the phase diagram is the remaining
impurity of the two separated phases even when the system undergoes an
infinite pressure. This reflects in the fact that the coexistence
lines end up at values of $x$ other than 1 or 0 (pure components) when
$\eta=1$ (actually, this effect is noticeable only for the values of the
large-cube rich phase, although it is also present in the other phase).
The prominent asymmetry of the coexistence line is another striking
feature, but easy to understand: it arises from the enormous volume
difference between large and small cubes necessary to produce demixing
(notice
that demixing begins for $r\approx 10$, and this means that large cubes
occupy a volume 1000 times larger than the small ones). This forces the 
large-cube impurities in the small-cube rich phase to be in an extremely
low concentration. The wide metastability region in the small-cube
rich side means that for those compositions the mixture is less
sensitive to variations in composition.

\section{Freezing of the binary mixture}
\label{freezing}

In order to check to which extent the demixing scenario found in
the previous section holds we have to determine whether the 
fluids are stable against spatial modulations at the coexisting
compositions. Spatial inhomogeneities cause a divergence of the
structure factor matrix (\ref{struct}) at a certain nonzero wavevector.
Thus for a given composition, $x$, the spatial instability is found
as the lowest total packing fraction at which the determinant
\begin{equation}
{\cal M}({\bf k})\equiv\left|{\sf P}^{-1}-\hat{\sf C}({\bf k})\right|
\label{spatialinst}
\end{equation}
vanishes for at least one vector ${\bf k}$. We can use the expression
for the DCF found in I [Eqs.\ (53), (54), and (57) of I] and simplify
the problem by simply looking for instabilities along the three 
coordinate axes. By symmetry, this amounts to take ${\bf k}=(k,0,0)$.

\begin{figure}
\epsfig{file=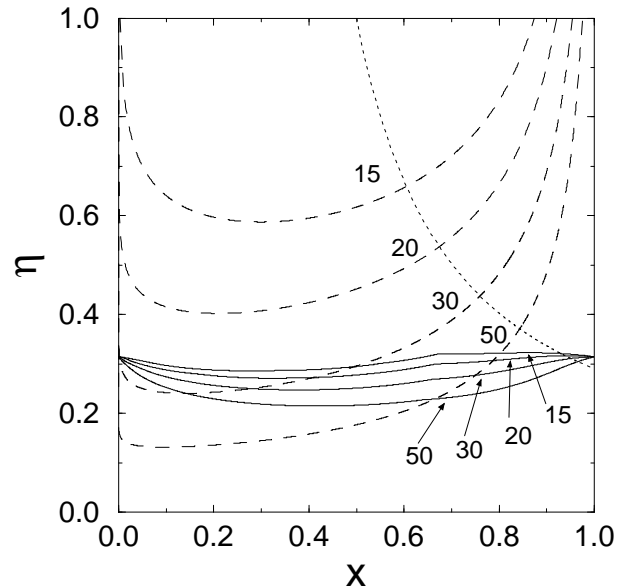, width=3.2in}
\caption[]{Phase diagram of the demixing transition of parallel hard cubes;
$\eta$ is the fraction of volume occupied by all cubes, whereas
$x\equiv\eta_1/\eta$ is the fraction of volume occupied by the large cubes.
Dashed lines represent the coexistence lines of the demixing transition 
for different values of $r=\sigma_1/\sigma_2$; solid lines are the
freezing spinodals for the same values of $r$;
finally, the dotted line is the line of critical points for all $r>r_c$.}
\label{solid}
\end{figure}

%
%The values of $\eta$ and $d=2\pi/k$ at which ${\cal M}({\bf k})=0$,
%as functions of $x$, are plotted in Figs.\ \ref{solid} and \ref{spacing},
%respectively. In Fig.\ \ref{solid} these lines of instability are
%
The value of $\eta$ at which ${\cal M}({\bf k})=0$,
as a function of $x$, is plotted in Fig.\ \ref{solid} for different
values of $r$; these lines
of instability are
compared with the coexistence lines of the demixing transition (Fig.\ 
\ref{demix}). It is clear from the figure that the critical points of
the latter are always in the unstable region; therefore, for any
pair of coexisting fluids, at least one (the large-cube rich one) is
always unstable against spatial inhomogeneities. In other words, of the
two phases in which the system phase separates, the one rich in large
cubes must always be a solid. Notice, on the other hand, that the other
phase is also unstable for size ratios smaller than $r\approx 30$. 
One is then tempted to conclude that fluid-fluid demixing is preempted
by freezing in this system. However coexistence between a
large-cube rich solid phase and a small-cube rich fluid phase may
change drastically the compositions of the coexisting phases and thus
make a fluid-solid demixing more stable than just a freezing of the
whole system. The only conclusion we can draw from Fig.\ \ref{solid}
is that the fluid-fluid demixing transition found in Sec.\ \ref{stability}
is always metastable.

%\begin{figure}
%\epsfig{file=fig4.eps, width=3.2in}
%\caption[]{Lattice constant, $d$ (in units of the large particle size
%$\sigma_1$), as a function of the relative
%%volume fraction of the large component, $x=\eta_1/\eta$, 
%at the solid spinodal, for four different size ratios. The curves shown
%in the main figure correspond to a crystal made of small cubes,
%and those of the inset to a crystal made of large cubes. The jumps 
%may indicate special lattice arrangements.}
%\label{spacing}
%\end{figure}

\section{Infinitely asymmetric binary mixture: parallel adhesive
hard cubes}
\label{adhesive}

\subsection{The binary mixture as an effective one-component fluid}
\label{effective}

In order to study the phase behaviour of a very asymmetric binary
mixture let us first consider the effect on the interaction of
the big particles induced by the small ones (depletion). To this
purpose let us use a semi-grand ensemble in which the
small particles (solvent) are kept at constant chemical potential.
This is the usual experimental setup in colloidal suspensions.
In this situation the element of the structure-factor matrix
(\ref{struct})
corresponding to the correlations between large particles can be
considered as the structure factor of an equivalent one-component
fluid made of the large particles interacting via the effective
potential induced by the solvent. This turns out to be a very useful
viewpoint. Let us see how this come about.

The appropriate thermodynamic potential for the semi-grand
ensemble is obtained through a Legendre transformation of the 
Helmholtz free energy, namely
\begin{eqnarray}
\Upsilon(\mu_2,[\rho_1])&=&F[\rho_1,\rho_2]-\mu_2\int d{\bf r}\,
\rho_2({\bf r}) \, , \label{upsilon} \\
\mu_2&=&\frac{\delta F}{\delta\rho_2({\bf r})} =
\beta^{-1}\ln{\cal V}_2\rho_2({\bf r})+
\frac{\delta F^{\rm ex}}{\delta\rho_2({\bf r})} \, ,
\label{rho2equil}
\end{eqnarray}
where Eq.\ (\ref{rho2equil}) provides the equilibrium density of
the solvent for a given chemical potential $\mu_2$ and a solute
density profile $\rho_1({\bf r})$, thus allowing us to eliminate
$\rho_2({\bf r})$ from the r.h.s.\ of (\ref{upsilon}). The
thermodynamic potential $\Upsilon$ can also be looked at as the
{\em Helmholtz free energy} functional of an effective one-component
fluid, for which $\mu_2$ is just an external parameter tuning the
interaction between its particles. Accordingly we can separate out
the ideal and excess parts,
\begin{eqnarray}
\Upsilon(\mu_2,[\rho_1])&=&\beta^{-1}\int d{\bf r}\,\rho_1({\bf r})
\left[\ln{\cal V}_1\rho_1({\bf r})-1\right]  \nonumber \\
&+&\Upsilon^{\rm ex}(\mu_2,[\rho_1]) \, ,
\label{upsidupsex}
\end{eqnarray}
where, upon comparison with (\ref{upsilon}), 
\begin{eqnarray}
\Upsilon^{\rm ex}(\mu_2,[\rho_1])&\equiv&\beta^{-1}
\int d{\bf r}\,\rho_2({\bf r})
\left[\ln{\cal V}_2\rho_2({\bf r})-1\right] \nonumber \\
 &+&F^{\rm ex}[\rho_1,\rho_2]-\mu_2\int d{\bf r}\,\rho_2({\bf r}) \, ,
\label{upsex}
\end{eqnarray}
with $\rho_2({\bf r})$ determined by Eq.\ (\ref{rho2equil}). 

Now, the DCF of the effective fluid will be
\begin{equation}
c_{\rm eff}({\bf r},{\bf r}')=-\beta\frac{\delta^2\Upsilon^{\rm ex}}
{\delta\rho_1({\bf r})\delta\rho_1({\bf r}')} \, .
\label{effDCFdef}
\end{equation}
The first functional derivative of $\Upsilon^{\rm ex}$ can be written
\begin{eqnarray}
\frac{\delta\Upsilon^{\rm ex}}{\delta\rho_1({\bf r})} &=& \int
d{\bf s}\,
\left\{\beta^{-1}\ln{\cal V}_2\rho_2({\bf s})+
\frac{\delta F^{\rm ex}}{\delta\rho_2({\bf s})}-\mu_2\right\}
\frac{\delta\rho_2({\bf s})}{\delta\rho_1({\bf r})} \nonumber \\
&+&\frac{\delta F^{\rm ex}}{\delta\rho_1({\bf r})} =
\frac{\delta F^{\rm ex}}{\delta\rho_1({\bf r})} \, ,
\label{dupsilon}
\end{eqnarray}
A new derivative yields
\begin{equation}
c_{\rm eff}({\bf r},{\bf r}')=c_{11}({\bf r},{\bf r}')+
\int d{\bf s}\,A({\bf r},{\bf s})c_{21}({\bf s},{\bf r}') \, ,
\label{Ceff}
\end{equation}
where we have used (\ref{effDCFdef}), the usual DCF matrix definition
$c_{ij}({\bf r},{\bf s})=-\beta\delta^2F^{\rm ex}/\delta\rho_i({\bf r})
\delta\rho_j({\bf s})$, and the shorthand
$A({\bf r},{\bf s})\equiv\delta\rho_2({\bf s})/
\delta\rho_1({\bf r})$.

The functional $A({\bf r},{\bf s})$ can be readily obtained by
deriving (\ref{rho2equil}) with respect
to $\rho_1({\bf r})$, which leads to
\begin{equation}
\frac{1}{\rho_2({\bf s})}A({\bf r},{\bf s})-c_{12}({\bf r},{\bf s})-
\int d{\bf t}\,A({\bf r},{\bf t})c_{22}({\bf t},{\bf s})=0 \, .
\label{Aeq}
\end{equation}
Then
\begin{equation}
A({\bf r},{\bf s})=\int d{\bf t}\,c_{12}({\bf r},{\bf t})
B({\bf t},{\bf s}) \, ,
\label{AinvB}
\end{equation}
where $B$ is the solution to
\begin{equation}
\int d{\bf t}\,\left\{\frac{1}{\rho_2({\bf r})}
\delta({\bf r}-{\bf t})-c_{22}({\bf r},{\bf t})\right\}
B({\bf t},{\bf s})=\delta({\bf r}-{\bf s}) \, .
\label{Bdef}
\end{equation}
Substitution of (\ref{AinvB}) into (\ref{Ceff}) finally leads to
%\begin{eqnarray}
%c_{\rm eff}({\bf r},{\bf r}')&=&c_{11}({\bf r},{\bf r}') \nonumber \\
%&+&\iint d{\bf t}d{\bf s}\,c_{12}({\bf r},{\bf t})B({\bf t},{\bf s})
%c_{21}({\bf s},{\bf r}') \, .
%\label{CeffB}
%\end{eqnarray}
\begin{equation}
c_{\rm eff}({\bf r},{\bf r}')=c_{11}({\bf r},{\bf r}')
+\iint d{\bf t}d{\bf s}\,c_{12}({\bf r},{\bf t})B({\bf t},{\bf s})
c_{21}({\bf s},{\bf r}') \, .
\label{CeffB}
\end{equation}

This expressions have full generality: it is valid for any binary
mixture of any kind and in any particular phase. But we can get a
bit further
if we particularise for the uniform fluid. In this case the
density profiles are constant and all the above functions depend on
the difference of their arguments,
because of translational invariance. Then, a Fourier transform
of Eq.\ (\ref{Bdef}) permits to obtain explicitly $B$ as
\begin{equation}
\widehat{B}({\bf k})=\frac{\rho_2}{1-\rho_2\hat{c}_{22}({\bf k})} \, .
\label{Bfour}
\end{equation}
Again Fourier transforming (\ref{CeffB}) and using (\ref{Bfour}) finally
yields
\begin{equation}
\hat{c}_{\rm eff}({\bf k})=\hat{c}_{11}({\bf k})+
\frac{\rho_2\hat{c}_{12}({\bf k})^2}{1-\rho_2\hat{c}_{22}({\bf k})} \, .
\label{Cefffour}
\end{equation}

In order to understand the meaning of Eq.\ (\ref{Cefffour}) let
us compute the structure factor of the effective fluid:
\begin{equation}
\rho_1S_{\rm eff}({\bf k})=\frac{1}{\rho_1^{-1}-
\hat{c}_{\rm eff}({\bf k})}=\frac{\rho_2^{-1}-\hat{c}_{22}({\bf k})}
{\left|{\sf P}^{-1}-\hat{\sf C}({\bf k})\right|} \, ;
\label{structeff}
\end{equation}
the resulting expression is but the $(1,1)$-element of the
structure-factor matrix 
$\rho\widehat{\sf S}({\bf k})$, according to its definition
(\ref{struct}). This was, by the way, the starting point from which
the effective fluid was defined in Ref.\ \onlinecite{heno}.

Further insight can be gained if we expand the second term of
(\ref{Cefffour}) in powers of $\rho_2$:
\begin{equation}
\hat{c}_{\rm eff}({\bf k})=\hat{c}_{11}({\bf k})+
\rho_2\sum_{n=0}^{\infty}\rho_2^n\hat{c}_{12}({\bf k})
[\hat{c}_{22}({\bf k})]^n\hat{c}_{21}({\bf k}) \, ,
\label{Ceffexpan}
\end{equation}
where it can be explicitly seen that while the first term in (\ref{Cefffour})
represent the direct correlation between two solute particles via the
direct potential between them, the second term accounts for the indirect
contributions to this correlation due to interaction with one, two, three,
etc, intermediate solvent particles. Is this effect that accounts
for depletion in the binary mixture of hard particles.

\subsection{Depletion in the binary mixture of parallel hard cubes}
\label{depletion}

\begin{figure}
\epsfig{file=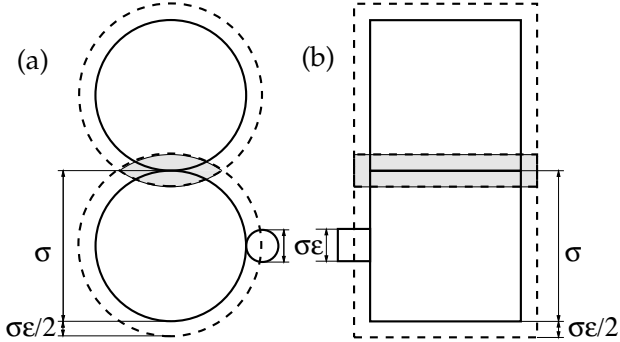, width=1.9in, angle=90}
\caption[]{Increment in the total volume volume, $\Delta V$, available
to the small particles
when two large particles come to touch each other (shaded region).
If $v_0$ denotes the volume of a large particle, $\Delta v$ is (a)
$v_0\epsilon^2(3/2+\epsilon)$ for HS, and (b) $v_0\epsilon(1+\epsilon^2)$, 
for PHC.}
\label{overlap}
\end{figure}

Let us first compare the effective attraction between large particles
induced by the small ones (depletion). A simple way to achieve this
is by computing the work we have to make against the system in order to
separate two big particles further than the diameter of a small one.
This work will be simply $P\Delta V$, with $P$ the pressure of the
fluid and $\Delta V$ the free volume lost by the small particles
due to the disappearance of the overlap between the excluded regions
of the large particles (shaded in Fig.\ \ref{overlap}).

In the case of HS $\Delta V=(3v_0/2)\epsilon^2+O(\epsilon^3)$, with
$v_0$ the volume of a big sphere and $\epsilon$ the small-to-large
diameter ratio. It means that in the diluted regime of the small
particles ($P\sim\rho_2$) this work can be estimated as
$\sim \beta^{-1}(3/2)\eta_2/\epsilon$.
In the case of PHC $\Delta V=v_0\epsilon+O(\epsilon^2)$, with
$v_0$ the volume of a big cube and $\epsilon$ the small-to-large
edge-length ratio. Again in the diluted regime of the small cubes
the work is $\sim \beta^{-1}\eta_2/\epsilon^2$.

In other words, the depletion induced by PHC is much stronger
than that induced by HS. In the infinite asymmetry limit, the binary
mixture HS has been shown to reduce to the fluid of adhesive HS,
provided $\eta_2$ is kept constant.\cite{heno} According to our
estimation, in order to have a similar limit for the binary mixture
of PHC we must scale the packing fraction of the small cubes as
$\eta_2=\epsilon\xi$, with $\xi$ a constant.

We can now assume this scaling of $\eta_2$ and take the $\epsilon\to 0$
limit in Eq.\ (\ref{Cefffour}). A tedious but straightforward calculation
leads to
\begin{eqnarray}
c_{\rm eff}({\bf r}) &=& c_{\rm PHC}({\bf r})+c_{\rm ad}({\bf r}) \, ,
\label{CeffCPHCCad} \\
c_{\rm ad}({\bf r}) &=& \frac{\xi}{2(1-\eta)^2}
\left\{\delta_{\rm S}({\bf r})+yS({\bf r})+6y^2V({\bf r})\right\} \, ,
\label{Cad}
\end{eqnarray}
where $y\equiv\eta_1/(1-\eta_1)$, $c_{\rm PHC}({\bf r})$ is the DCF of the
one-component PHC fluid [Eq.\ (56) of I], and
\begin{eqnarray}
\delta_{\rm S}({\bf r})&=& A(x,y,z)+A(y,z,x)+A(z,x,y)  \, ,
\label{deltacontact} \\
A(u,v,w) &\equiv& \delta(\sigma_1-|u|)L(v)L(w) \, , \label{contactarea} \\
L(u) &\equiv& (\sigma_1-|u|)\Theta(\sigma_1-|u|) \, , \label{L}
\end{eqnarray}
$\Theta(t)$ being the usual Heaviside step function. Equation
(\ref{deltacontact}) represents a delta function at contact of two
large cubes, multiplied by the contact surface. The functions $S({\bf r})$
and $V({\bf r})$ are the overlap surface and volume, respectively, which
already appear in the definition of $c_{\rm PHC}$ [Eq.\ (53) of I].
In the zero density limit $c_{\rm eff}$ becomes
\begin{equation}
c_{\rm eff}({\bf r})\sim f({\bf r})+\xi\delta_{\rm S}({\bf r}) \, ,
\label{Ceffzero}
\end{equation}
with $f({\bf r})$ the Mayer function of the large cubes;
so, as in HS, in the infinitely asymmetric mixture depletion induces an
adhesive potential (in this case, of strength $\xi$). We will henceforth
refer to this effective fluid as the fluid of {\em parallel adhesive
hard cubes} (PAHC).

\subsection{Free energy functional of the fluid of parallel adhesive hard
cubes}
\label{PAHCfluid}

Let us now take the $\epsilon\to 0$ limit in the functional
(\ref{upsilon}) to obtain the Helmholtz free-energy functional for
the fluid of PAHC. In the limit we will find that $\Upsilon\to\infty$;
however, this is not a problem as long as for {\em every} fixed $\epsilon$
there is a well-defined functional giving rise to a phase behaviour
which do have a finite limit when $\epsilon\to 0$. We will show that
this is the case, and thus this functional will be the effective
functional we are looking for. We will see that the infinite contribution
is just a {\em constant} shift in the origin of free energies,
absolutely irrelevant for the phase behaviour.

Let us begin by recalling what FMT prescribes for the semi-grand
potential (\ref{upsilon}). It is convenient to introduce two dimensionless
densities, $\eta({\bf r})\equiv\sigma_1^3\rho_1({\bf r})$, and 
$\xi({\bf r})\equiv\sigma_1^3\epsilon^2\rho_2({\bf r})$. [There is
no possible confusion between the function $\eta({\bf r})$
and the total packing fraction, because when $\epsilon\to 0$ the total
packing fraction is simply the packing fraction of the large component,
i.e.\ the average of $\eta({\bf r})$.] In what follows we will fix
the unit length of our system by choosing $\sigma_1=1$. In terms of
these functions the FMT form of the functional (\ref{upsilon}) is
\begin{eqnarray}
\beta\Upsilon&=&\beta\overline{F}^{\rm id}+\int d{\bf r}\,\Phi({\bf r})
\nonumber \\
&&+\epsilon^{-2}\int d{\bf r}\,\xi({\bf r})\Bigl(\ln[{\cal V}_2
\epsilon^{-2}\xi({\bf r})]-1-\beta\mu_2\Bigr) \, ,  \label{upsFMT} \\
\beta\overline{F}^{\rm id}&=&\int d{\bf r}\,\eta({\bf r})
\Bigl(\ln[{\cal V}_1\eta({\bf r})]-1\Bigr) \, .
\label{PAHCideal}
\end{eqnarray}
Equation (\ref{PAHCideal}) is just the ideal Helmholtz free energy
of the effective fluid, and $\Phi({\bf r})$ is given by (\ref{excess}),
where, in the current notation, $n_{\alpha}=\eta*\omega_1^{(\alpha)}+
\epsilon^{-2}\xi*\omega_2^{(\alpha)}$. But $\xi({\bf r})$ is a dependent
variable which should be eliminated in terms of $\eta({\bf r})$ and 
$\mu_2$ via Eq.\ (\ref{rho2equil}), which in our case reads
\begin{equation}
\ln\xi({\bf r})=\ln z-\sum_{\alpha}\frac{\partial\Phi}{\partial n_{\alpha}}
*\omega^{(\alpha)}({\bf r}) \, ,
\label{equilxi}
\end{equation}
where we have defined the renormalised fugacity $z\equiv\epsilon^2
\exp(\beta\mu_2^{\rm ex})$, $\mu_2^{\rm ex}\equiv\mu_2-\beta^{-1}
\ln{\cal V}_2$. For $\xi({\bf r})$ to have a well-defined expansion 
in powers of $\epsilon$ we are forced to assume that $z=O(1)$.

We are almost ready to carry out the $\epsilon$-expansion. It only remains
to determine the contributions to this expansion coming from convolutions
with $\omega_2^{(\alpha)}$. Let $f(u)$ be an arbitrary function of a
single variable $u$. Then
\begin{eqnarray}
f*\theta^u_2&=&\int_{u-\epsilon/2}^{u+\epsilon/2}f(t)\,dt=
\epsilon f(u)+O(\epsilon^3) \, , \label{ftheta}  \\
f*\delta^u_2&=&\frac{1}{2}[f(u+\epsilon/2)+f(u-\epsilon/2)] \nonumber \\
&&=f(u)+\frac{\epsilon^2}{8}f''(u)+O(\epsilon^4) \, . \label{fdelta}
\end{eqnarray}
Accordingly if $f({\bf r})$ is an arbitrary function of ${\bf r}$,
from the definitions (\ref{omegas}) and the expansions above it
follows 
\begin{mathletters}
\begin{eqnarray}
f*\omega^{(0)}_2 &=& f+(\epsilon^2/8)\nabla^2f+O(\epsilon^4) \, ,
\label{f0expan} \\
f*\omega^{(\alpha)}_2 &=& \epsilon^{\alpha}f+O(\epsilon^{\alpha+2}) \, ,
\label{faexpan}
\end{eqnarray}
\label{fexpan}
\end{mathletters}

\noindent
for $\alpha=3$ or any vector component of $\alpha=2$ and 1. Then, 
assuming $\xi=\xi_0+\xi_1\epsilon+\xi_2\epsilon^2+O(\epsilon^3)$, the
weighted densities $n_{\alpha}$ can be expanded as
\begin{mathletters}
\begin{eqnarray}
n_0 &=& \xi_0\epsilon^{-2}+\xi_1\epsilon^{-1}+\left\{\overline{n}_0+
\xi_2+\frac{1}{8}\nabla^2\xi_0\right\}+O(\epsilon) \, ,
\label{nexpan0}  \\
{\bf n}_1 &=& \xi_0{\bf u}\epsilon^{-1}+\left\{\xi_1{\bf u}+
\overline{\bf n}_1\right\}+O(\epsilon) \, ,
\label{nexpan1}  \\
{\bf n}_2 &=& \left\{\xi_0{\bf u}+\overline{\bf n}_2\right\}+
\xi_1{\bf u}\epsilon+O(\epsilon^2) \, ,
\label{nexpan2}  \\
n_3 &=& \overline{n}_3+\epsilon\xi_0+\epsilon^2\xi_1+
\epsilon^3\xi^2+O(\epsilon^4) \, , 
\label{nexpan3}
\end{eqnarray}
\label{nexpan}
\end{mathletters}

\noindent
where ${\bf u}\equiv(1,1,1)$ and $\overline{n}_{\alpha}\equiv\eta*
\omega_1^{(\alpha)}$. 
\end{multicols}
\widetext
\noindent\rule{3.375in}{0.1mm}

%%%%%%%%%%%%%% acabar widetext con %%%%%%%%%%%%%%%%%%%%%%%
%\hspace*{3.625in}\rule{3.375in}{0.1mm}
%\begin{multicols}{2}
%\narrowtext\noindent
%%%%%%%%%%%%%%%%%%%%%%%%%%%%%%%%%%%%%%%%%%%%%%%%%%%%%%%%%%
From the expansions (\ref{fexpan}), (\ref{nexpan}), we can obtain
\begin{eqnarray}
\sum_{\alpha}\frac{\partial\Phi}{\partial n_{\alpha}}*\omega^{(\alpha)}_2
&=& -\ln(1-\overline{n}_3)+\frac{8\xi_0+\overline{n}_2}{1-\overline{n}_3}
\epsilon     \nonumber \\
&&+\left\{ \frac{1}{8}\nabla\cdot\left(\frac{\nabla\overline{n}_3}
{1-\overline{n}_3}\right)+\frac{1}{2}\frac{\overline{n}_2^2-
\overline{\bf n}_2\cdot\overline{\bf n}_2}{(1-\overline{n}_3)^2}+
\frac{8\xi_1+\overline{n}_1}{1-\overline{n}_3}
+\frac{(27/2)\xi_0^2+
4\xi_0\overline{n}_2}{(1-\overline{n}_3)^2} \right\}\epsilon^2+
O(\epsilon^3) \, .
\label{sumalpha}
\end{eqnarray}
On the other hand,
$\ln\xi=\ln\xi_0+\frac{\xi_1}{\xi_0}\epsilon+\left\{\frac{\xi_2}{\xi_0}
-\frac{\xi_1^2}{2\xi_0^2}\right\}\epsilon^2+O(\epsilon^3)$, hence
(\ref{equilxi}) implies
\begin{mathletters}
\begin{eqnarray}
\xi_0 &=& z(1-\overline{n}_3)  \, ,  \label{xi0} \\
\xi_1 &=& -z\overline{n}_2-8z^2(1-\overline{n}_3) \, , \label{xi1} \\
\xi_2 &=& -\frac{z}{8}\nabla^2\overline{n}_3-z\overline{n}_1
+12z^2\overline{n}_2+\frac{37}{2}z^3(1-\overline{n}_3)
-\Phi^{\rm ad} \, , \label{xi2}
\end{eqnarray}
\label{xiexpan}
\end{mathletters}

\noindent
where for convenience we have introduced the shorthand
\begin{equation}
\Phi^{\rm ad}\equiv
\frac{z}{8}\frac{|\nabla\overline{n}_3|^2-4\overline{\bf n}_2\cdot
\overline{\bf n}_2}{1-\overline{n}_3}   \, . 
\label{Phiad}
\end{equation}

We have already explicitly eliminated $\xi$ in terms of $\eta$ and 
$z$, with the help of the expansion in $\epsilon$. We can now proceed
to expand $\Upsilon$ itself, but before, let us rewrite the excess 
part appearing in (\ref{upsFMT}) as
\begin{equation}
\beta\Upsilon^{\rm ex}=\epsilon^{-2}\int\xi\left[\ln\left(\frac{\xi}
{z(1-\overline{n}_3)}\right)-1\right]
+\int\Bigl\{\Phi+\epsilon^{-2}\xi\ln(1-\overline{n}_3)\Bigr\} \, ;
\label{upsexcess}
\end{equation}
then, using (\ref{nexpan}) and (\ref{xiexpan}), and defining
$\overline{\Phi}$ as in (\ref{excess}) with the $n_{\alpha}$ replaced
by $\overline{n}_{\alpha}$ (i.e., the excess free-energy functional
of the one-component PHC fluid),
\begin{eqnarray}
\Phi+\epsilon^{-2}\xi\ln(1-\overline{n}_3) &=&
\Bigl\{z\overline{n}_2+4z^2(1-\overline{n}_3)\Bigr\}\epsilon^{-1}
\label{term1}  \\
&+&\Biggl\{z\overline{n}_1
-14z^2\overline{n}_2-\frac{119}{2}z^3(1-\overline{n}_3)-\frac{z}{2}
\frac{\overline{n}_2^2}{1-\overline{n}_3}
+\frac{z}{8}\nabla\cdot\Bigl[\ln(1-\overline{n}_3)\nabla\overline{n}_3
\Bigr]+\overline{\Phi}+\Phi^{\rm ad}\Biggr\}+O(\epsilon) \, , 
\nonumber \\
\xi\ln\left(\frac{\xi}{z(1-\overline{n}_3)}\right)-\xi &=&
-z(1-\overline{n}_3)
+\left\{8z^2\overline{n}_2+32z^3(1-\overline{n}_3)
+\frac{z}{2}\frac{\overline{n}_2^2}{1-\overline{n}_3}\right\}\epsilon^2
+O(\epsilon^3) \, .
\label{term2}
\end{eqnarray}
\hspace*{3.625in}\rule{3.375in}{0.1mm}
\begin{multicols}{2}
\narrowtext\noindent
Substituting these two expansions into (\ref{upsexcess}), and using
$\int\nabla\cdot[\ln(1-\overline{n}_3)\nabla\overline{n}_3]=0$, which
holds if the density is constant at the boundaries or
if it is a periodic function, we finally obtain
\begin{equation}
\Upsilon=-\Pi_0(\epsilon)V+\mu_0(\epsilon)N+
\overline{F}+F^{\rm ad}+O(\epsilon) \, ,
\label{finUpsilon}
\end{equation}
where $\overline{F}$ is the FMT free-energy functional of the
fluid of the large PHC; $V$ and $N$ are, respectively, the system
volume and the number of large cubes; $F^{\rm ad}=\int\Phi^{\rm ad}$
is the new adhesive term; and $\Pi_0$ and $\mu_0$ are given by
\begin{eqnarray}
\beta\Pi_0(\epsilon) &=& z\epsilon^{-2}-4z^2\epsilon^{-1}+
\frac{55}{2}z^3+O(\epsilon) \, , \label{Pi0}  \\
\beta\mu_0(\epsilon) &=& z\epsilon^{-2}+\{3z-4z^2\}\epsilon^{-1}
\nonumber \\
&&+\left\{\frac{55}{2}z^3-18z^2+3\right\}+O(\epsilon) \, . \label{mu0}
\end{eqnarray}

The term $-\Pi_0(\epsilon)V+\mu_0(\epsilon)N$ is divergent
with $\epsilon\to 0$. It is
the contribution of the small cubes to the free energy (as a
matter of fact, their density is infinite in this limit).
However it is irrelevant for the phase
behaviour of the effective fluid because it simply adds
$\Pi_0$ to the pressure and $\mu_0$
to the chemical potential; as these two terms are independent of 
the density, they just cancel out in the equilibrium equations.
Accordingly the final free energy functional for the effective
one-component PAHC fluid turns out to be
\begin{equation}
F_{\rm PAHC}([\rho];z)=F_{\rm PHC}[\rho]+F^{\rm ad}([\rho];z) \, .
\label{freePAHC}
\end{equation}
As a selfconsistency test, it is straightforward to show
[using Eq.\ (\ref{xi0})] that $c^{\rm ad}({\bf r}-{\bf r}')=
-\beta\delta^2F^{\rm ad}/\delta\rho({\bf r})\delta\rho({\bf r}')$,
for the $c^{\rm ad}$ function defined in (\ref{Cad}).

\subsection{Phase behaviour of the infinitely asymmetric binary
mixture}
\label{phasebehaviour}

The phase behaviour of a very asymmetric binary mixture of PHC
can be understood from that of the effective fluid of PAHC, whose
FMT free-energy functional we have just derived.

\begin{figure}
\epsfig{file=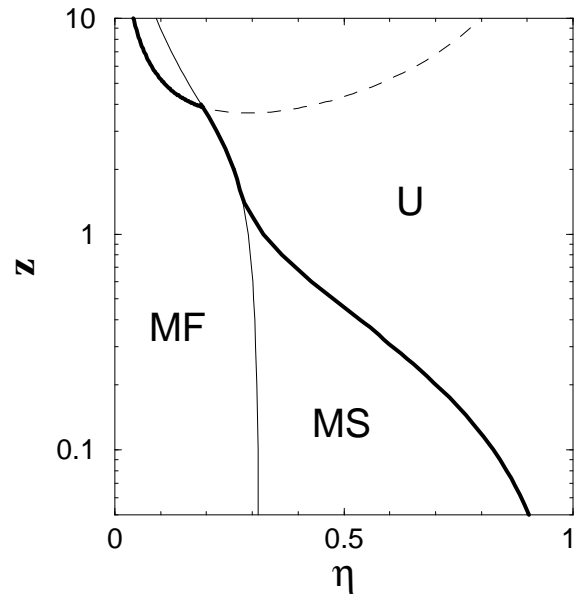, width=3.0in}
\caption{Solvent fugacity, $z$, vs.\ solute packing fraction, $\eta$,
of the infinitely asymmetric binary mixture of PHC.
The thick solid line separates the unstable region (U) from the
metastable one; the thin one marks the (continuous) transition
from a metastable fluid (MF) to a metastable solid (MS); the dashed
one is the fluid-fluid spinodal.}
\label{metastable}
\end{figure}

As concerns the phase behaviour of the uniform PAHC fluid, from
(\ref{freePAHC}), (\ref{excess}), and (\ref{Phiad}) we can readily
obtain the free energy per unit volume,
\begin{equation}
\beta f=\eta\left\{\ln{\cal V}_1-1+\ln y+3(1-z/2)y+y^2\right\} \, .
\label{uniformf}
\end{equation}
The pressure, $P=-\partial F/\partial V=y^2\partial(f/\eta)/\partial y$,
turns out to be
\begin{equation}
\beta P=y+3(1-z/2)y^2+2y^3 \, .
\label{uniformP}
\end{equation}
This equation has a van der Waals loop; the critical point can be
found as the solution of the equations $\partial P/\partial\eta=0$
and $\partial^2P/\partial\eta^2=0$, \cite{callen} i.e.
\begin{mathletters}
\begin{eqnarray}
1+6(1-z_c/2)y_c+6y_c^2 &=& 0  \, ,  \\
1-z_c/2+2y_c &=& 0 \, ,
\label{critical}
\end{eqnarray}
\end{mathletters}

\noindent
which is $z_c=2(1+\sqrt{2/3})\approx 3.63$, and $y_c=1/\sqrt{6}$,
i.e.\ $\eta_c=1/(1+\sqrt{6})\approx 0.29$.
On the other hand, the equation of the spinodal ($\partial
P/\partial\eta=0$) of this vapor-liquid transition is 
\begin{equation}
z=\frac{1+4\eta+\eta^2}{3\eta(1-\eta)} \, ;
\label{spinodalFF}
\end{equation}
it is plotted in Fig.\ \ref{metastable}.
Notice that this spinodal could have been obtained directly from
(\ref{det-3D}) by taking the limit $r\to\infty$, $x\to 1$, under
the constraint $r\eta(1-x)=\xi\to z(1-\eta)$ [the limit follows
from (\ref{xi0})]. Thus it is not surprising that
the line of critical points in Fig.\ \ref{demix} reaches $\eta_c$
for $x\to 1$. This makes clear the double interpretation
of this transition: as a vapor-liquid transition of the PAHC fluid
(with $z^{-1}$ playing the role of a temperature), or as a demixing
transition of the infinitely asymmetric binary mixture.

Freezing of this system into a simple cubic lattice is again a
continuous transition. Hence the transition line can be determine
by the procedure described in Sec.\ \ref{one-component}, i.e.\ as
the divergence of the structure factor [Eq.\ (\ref{structure})] with
the effective DCF (\ref{CeffCPHCCad}). The result is the line
shown in Fig.\ \ref{metastable}. As it occurred
for the general binary mixture
(see Sec.\ \ref{stability}), the freezing line crosses the demixing
spinodal at a packing fraction smaller than $\eta_c$; in other
words fluid-fluid demixing is a metastable transition.

So far we have gone no further than we did in Secs.\
\ref{stability}, \ref{freezing}. However this time we can study
fluid-solid coexistence because the density profile of the solvent
is absent from the description. To proceed we again parametrise 
the density of the large cubes as in (\ref{densprof}), also with
a prefactor $\vartheta$ to account for vacancies. We recall that
the lattice parameter is related to this occupancy ratio by
$d=(\vartheta/\eta)^{1/3}$. The ideal contribution to the free 
energy per particle, $\Psi$, is again given by (\ref{FNid})
(adding $\ln\vartheta$ from the vacancies), and the hard-core
part of the excess contribution is given by (\ref{FNex}) [with
$n_3({\bf r})=\vartheta p(x)p(y)p(z)$]. We now need to work
out the adhesive term. To this purpose first notice that
\begin{eqnarray}
p'(u)^2-q(u)^2 &=& -4\sum_{n,m=-\infty}^{\infty}g(u-nd+1/2)
\nonumber \\
&&\times g(u-md-1/2) \, ;
\label{simplify}
\end{eqnarray}
hence the adhesive free energy per particle can be written
\begin{eqnarray}
\Psi^{\rm ad} &=& -\frac{3}{2}\vartheta z
\sum_{n,m=-\infty}^{\infty}\int_{-d/2}^{d/2}dx   \nonumber \\
&&\times g(x-nd+1/2)g(x-md-1/2)U(x)   \nonumber \\
&=& -\frac{3}{2}\vartheta z\int_{-\infty}^{\infty}dx\,g(x) \nonumber \\
&&\times \left[\sum_{n=-\infty}^{\infty}g(x-nd+1)\right]U(x+1/2) \, ,
\label{FNad}
\end{eqnarray}
where $U(x)$ is defined as
\begin{equation}
U(x)\equiv\int_{-d/2}^{d/2}dy\int_{-d/2}^{d/2}dz
\frac{p(y)^2p(z)^2}{1-\vartheta p(x)p(y)p(z)}  \, ,
\label{U}
\end{equation}
and it is periodic with period $d$ [we have made use of periodicity
in obtaining (\ref{FNad})].

It is convenient to rewrite Eq.\ (\ref{U}) integrating by part with
respect to both variables, $y$ and $z$; in doing so this equation
becomes
\begin{eqnarray}
U(x) &=& \int_{-d/2}^{d/2}dy\,p'(y)\int_{-d/2}^{d/2}dz\,p'(z)V(x,y,z)
\, ,  \label{UofV}  \\
V({\bf r}) &\equiv& T(y)p(y)T(z)p(z)\frac{4-3n_3({\bf r})+
3n_3({\bf r})^2}{1-n_3({\bf r})} \, ,
\label{V}
\end{eqnarray}
where $T(u)=u$ if $u\in(-d/2,d/2)$ [the only relevant interval
in (\ref{UofV})] and it is $d$-periodic. Function $V(x,y,z)$ is
then also $d$-periodic in all its three arguments; accordingly
Eq.\ (\ref{UofV}) can be rewritten as
\end{multicols}
\widetext
\noindent\rule{3.375in}{0.1mm}
\begin{equation}
U(x)=\int_{-\infty}^{\infty}dy\,g(y)\int_{-\infty}^{\infty}dz\,g(z)
\left[V(x,y+1/2,z+1/2)-2V(x,y+1/2,z-1/2)+V(x,y-1/2,z-1/2)\right] \, ,
\label{finalU}
\end{equation}
and therefore
\begin{eqnarray}
\Psi^{\rm ad} &=&
-\frac{3}{2}\vartheta z\int_{-\infty}^{\infty}dx\,g(x)
\int_{-\infty}^{\infty}dy\,g(y)\int_{-\infty}^{\infty}dz\,g(z)
\left[\sum_{n=-\infty}^{\infty}g(x-nd+1)\right] \nonumber \\
&&\times \left[V(x+1/2,y+1/2,z+1/2)-2V(x+1/2,y+1/2,z-1/2)
+V(x+1/2,y-1/2,z-1/2)\right] \, ,
\label{FNadlast}
\end{eqnarray}
\hspace*{3.625in}\rule{3.375in}{0.1mm}
\begin{multicols}{2}
\narrowtext\noindent
also suitable for Gauss-Hermite numerical integration.

In order to understand the effect of the adhesive contribution
(\ref{FNad}) let us see its asymptotic behaviour when $\alpha\to\infty$
and $d\to 1^+$ (equivalently $\vartheta\to\eta^+$), a limit
which would represent a close packed solid. From its definition
(\ref{functionp}) $p(u)\sim 1$ in this limit; thus $U(x)\sim 1/(1-\eta)$.
On the other hand, $g(u)$ is very sharply peaked, so
\[  \int_{-\infty}^{\infty}dx\,g(x)\sum_{n=-\infty}^{\infty}g(x-nd+1)
\sim g(0)=\left(\frac{\alpha}{\pi}\right)^{1/2} \, , \]
therefore
\begin{equation}
\Psi^{\rm ad}\sim -\frac{3}{2}z\frac{\eta}{1-\eta}\left(
\frac{\alpha}{\pi}\right)^{1/2}\to -\infty \, .
\label{asymp}
\end{equation}
On the other hand $\Psi^{\rm ex}=O(1)$ in this limit, while
$\Psi^{\rm id}\sim 3\ln g(0)\sim(3/2)\ln\alpha$.
In other words, the total free energy per particle of the effective
fluid monotonically decreases as the system approaches the close
packing, regardless the value of density and solvent fugacity. This
means that the system always collapses, i.e.\ the equilibrium phase
behaviour is always a close-packed solid coexisting with an 
infinitely diluted gas. This singular phase diagram is not exclusive
of PAHC. For adhesive HS, the adhesiveness vs.\ packing fraction phase
diagram ($z$ plays the role of adhesiveness for PAHC) has recently
been mapped out from simulations of the square-well fluid in
the limit of narrow and deep wells. \cite{bolhuis} These simulations
prove that the only stable phases of this system are also a 
close-packed solid and an infinitely diluted gas. The reason for 
this pathology was put forward some years ago by Stell, \cite{stell}
who showed that the partition function of adhesive HS diverges if
the number of particles is $N\geq 12$ (precisely the coordination
number of the fcc solid lattice).

In spite of the above we have seen that, as a function of $\alpha$
and $\vartheta$, the free energy per particle exhibits local
minima at any value of $z$ for some range of densities, the
smaller $z$ the wider this range. These local minima correspond
to metastable phases. The upper bound to the packing fractions
at which local minima exist for a given $z$ can be determined as
the point where the compressibility vanishes. This upper bound, as
a function of $z$, appears in Fig.\ \ref{metastable}. This
figure shows the metastable phases. Notice that the region of 
metastability widens as $z$ decreases. At low $z$ these metastable
phases are separated from the ``collapse'' by a large free energy
barrier, so the system spends a long time in them before
eventually becoming a close-packed solid. As a matter of fact,
if the system is prepared as a metastable solid at low $z$, for
a long time it will show a pseudo-coexistence between two
solid phases (an expanded solid and a close-packed solid). The
situation resembles the isostructural solid-solid transition
reported to occur in some colloidal fluids with a narrow and
deep attractive well. \cite{bolhuis,tejero} At higher $z$ the
same pseudo-coexistence should be observed between a diluted
fluid and a close-packed solid.

It is interesting to compare this phase behaviour with what 
has been determined to occur for adhesive HS using an 
effective-liquid DFT. \cite{tejero2} Fluid and solid also 
appear as local
minima of the free energy per particle as a function of the
gaussian width; however, above a certain line the free energy
becomes concave up to close packing. This puzzling behaviour
was interpreted in Ref.\ \onlinecite{tejero2} as a percolation
transition. In the light of our findings it is the equivalent to
the instability line of Fig.\ \ref{metastable}. The fluid of
adhesive HS also collapses into a close-packed solid. \cite{bolhuis}
The reason why this collapse has not been observed in Ref.\
\onlinecite{tejero2} is that the theory used there does not
account for vacancies, and this forces the lattice parameter
to be larger than 1 at {\em any} packing fraction. The
instability manifests itself as the reported loss of convexity
of the free energy.

\subsection{Polydispersity in the large cubes}
\label{polydispersity}

The singularity of the adhesive potential can be avoided by
introducing polydispersity in the size of the particles.
\cite{stell,frenkel} It is clear that this prevents the system
to form a perfectly packed solid. To see this effect on the
binary mixture we have introduced a small amount of polydispersity
in the size of the large cubes. It is very easy to realise
that starting off from a mixture of polydisperse large cubes
and small cubes and repeating the process described in Sec.\
\ref{PAHCfluid} we end up with exactly the same form of the
functional (\ref{freePAHC}), with the $\overline{n}_{\alpha}$'s now
replaced by those corresponding to the polydisperse mixture.

In order to make the simplest choice
we consider the cubes as parallelepipeds and choose the length
of each axis independently from a gaussian distribution of mean 1
and variance $\Delta\sigma$. This particular choice has two
important advantages (they will be made clear below): (i) the
free energy of the fluid phase is the same as that of the
monodisperse system (hence its phase behaviour as well), and
(ii) formally the expressions for the free energy of the solid 
phase change very little. It also has two drawbacks: (i) particles
are not cubic anymore, and (ii) there is a nonzero contribution
in the negative lengths. As these two inconvenients disappear when
$\Delta\sigma\to 0$ they can be overcome by choosing 
$\Delta\sigma\ll 1$. This choice also allows us to make two
more simplifying assumptions: 
(i) the ordered phase must be a substitutional solid, i.e.\ the
density profile can be expressed as
$\rho({\bf r}){\cal P}(\boldsymbol{\sigma})$,
with ${\cal P}$ the normalised size distribution,
$\boldsymbol{\sigma}\equiv(\sigma_x,\sigma_y,\sigma_z)$, and (ii)
phase separation induced by polydispersity \cite{sear} can be ignored.

Then, according to the definition of the $\overline{n}_{\alpha}$'s
\begin{equation}
\overline{n}_{\alpha}({\bf r}) = \int d\boldsymbol{\sigma}
{\cal P}(\boldsymbol{\sigma})\rho*
\omega^{(\alpha)}_{\boldsymbol{\sigma}}({\bf r})
= \rho*\widetilde{\omega}^{(\alpha)}({\bf r}) \, ,
\label{newnalpha}
\end{equation}
i.e.\ it has the same definition as in the monodisperse case, but
the weights are redefined as 
$\widetilde{\omega}^{(\alpha)}({\bf r})\equiv
\int d\boldsymbol{\sigma}{\cal P}(\boldsymbol{\sigma})
\omega^{(\alpha)}_{\boldsymbol{\sigma}}({\bf r})$.
This amounts to replacing $\theta^u$ and $\delta^u$ in
(\ref{omegas}) by 
\begin{mathletters}
\begin{eqnarray}
\tilde\theta^u &=& \frac{1}{2}\left[1-\mbox{erf}\,\left(\sqrt{2}
\frac{|u|-1/2}{\Delta\sigma}\right)\right] \, ,
\label{newtheta}  \\
\tilde\delta^u &=& \frac{1}{\sqrt{2\pi}\Delta\sigma}
\sum_{\{\pm\}}
\exp\left\{-2\frac{(u\pm 1/2)^2}{\Delta\sigma^2}\right\}  \, ,
\label{newdelta}
\end{eqnarray}
\label{newweights}
\end{mathletters}

\noindent
which are like smoothed counterparts of the original weights.
Since $\int_{-\infty}^{\infty}du\,\tilde\theta^u=
\int_{-\infty}^{\infty}du\,\tilde\delta^u=1$, it follows that
the free energy of the uniform fluid is the same as that of
the monodisperse system. Hence the fluid-fluid spinodal is 
the same as that shown in Fig.\ \ref{metastable}.

\begin{figure}
\epsfig{file=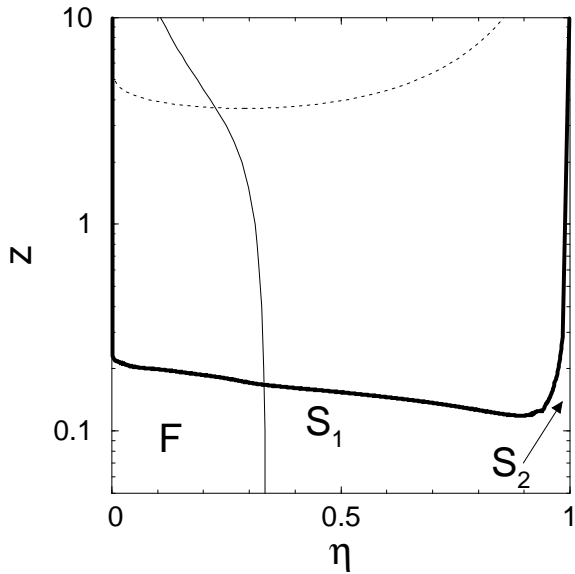, width=3.0in}
\caption{Same as Fig.\ \ref{metastable} for the polydisperse
PAHC fluid ($\Delta\sigma=0.045$).
The thick solid line marks the fluid-solid or solid-solid
coexistence; the thin one marks again the (continuous) fluid-solid
transition below the coexistence region; the dotted line is the
metastable fluid-fluid coexistence.}
\label{poly}
\end{figure}

We can determine the coexistence between the two fluid phases by 
means of the usual double tangent construction. \cite{callen}
Figure \ref{poly} shows the resulting coexistence line.

We can also assume a solid-like density profile as in the monodisperse
case. Surprisingly enough, in spite of the striking difference of
the smoothed weights defined by (\ref{newweights}) with respect to
the original ones, when we obtain the corresponding weighted densities
and work out the expressions a little bit, it turns out that the 
free-energy per particle of the polydisperse solid is simply given
by $\Psi=\Psi^{\rm poly}+\Psi^{\rm id}+\Psi^{\rm ex}+\Psi^{\rm ad}$,
where the last three contributions are given by Eqs.\ (\ref{FNid}),
(\ref{FNex}), and (\ref{FNadlast}), with the slight modification that
the parameter $\alpha$ appearing in the definitions (\ref{gaussians})
and (\ref{functione}) must be replaced by 
\begin{equation}
\widetilde{\alpha}=\frac{\alpha}{1+\alpha\Delta\sigma^2/2}
\label{newalpha}
\end{equation}
(of course, in these expressions $\sigma=1$, the mean value, and
$\Psi^{\rm id}$ carries the additional $\ln\vartheta$ to account
for vacancies), and where $\Psi^{\rm poly}=-\ln\left(\sqrt{2\pi}
\Delta\sigma\right)-1$ is the entropy of mixing (an irrelevant
constant).

From Eq.\ (\ref{newalpha}) it can be seen that no matter how small
$\Delta\sigma$ be, for small $\alpha$'s ($\alpha\ll\Delta\sigma^{-2}$)
$\widetilde{\alpha}\sim\alpha$, and the system is ``blind'' to 
polydispersity, whereas for large $\alpha$'s ($\alpha\gg
\Delta\sigma^{-2}$) $\widetilde{\alpha}\sim 2\Delta\sigma^{-2}$,
i.e.\ the system never collapses. 
As a consequence the singular behaviour of the monodisperse system
is removed, and we can readily determine phase
equilibria. A typical result for a small value of $\Delta\sigma$
is shown in Fig.\ \ref{poly}. This figure reveals several remarkable 
features. Firstly, it shows that the fluid-fluid transition is
metastable. Secondly, there is an isostructural solid-solid transition
for a certain interval of $z$. In this interval the expanded solid
(S$_1$) appears after a continuous transition from the fluid phase (F).
The expanded solid (S$_1$)-dense solid (S$_2$) transition ends at a
critical point, $z_s$, below which we can only find a fluid and a single
solid, separated by a continuous transition. Thirdly, as $z$ increases
from $z_s$ the expanded solid packing fraction decreases down to meeting
the freezing packing fraction. Above the point where this occurs 
coexistence is between a fluid and a dense solid (S$_2$), the former
quickly becoming highly diluted and the latter highly packed. Notice
the strong resemblance between this true equilibrium phase behaviour
and the metastable behaviour described to occur the monodisperse
PAHC fluid (Sec.\ \ref{phasebehaviour}).

\section{Discussion and conclusions}
\label{conclusions}

The fluid of PHC is a rather academic one which however has the
great advantage of being analytically tractable in contexts where
the fluid of HS is not, thanks to its adequacy to a fundamental
measure description. Yet, with some peculiarities due to the
lack of rotational symmetry, \cite{kirkpatrick,jagla} the physics
it reveals is similar to that of more realistic fluids. It then
allows for theoretical investigation on fluid phase behaviour otherwise
very difficult (the closely related fluid of parallel hard
parallelepipeds has been recently used in the context of associating
fluids \cite{cuestasear} and liquid crystals \cite{yurietal}).
The main contribution of the fluid of PHC is to the understanding
of the phase diagram of a binary mixture. This fluid proved to
undergo stronger depletion than HS. \cite{dijkstra,cuesta2} 
However, as it has been shown in this work, this feature is 
irrelevant when the effect of depletion in the solid phase is
accounted for. Spatial order of the large component strongly
enhances demixing, so that fluid-solid demixing becomes the main
scenario of the phase diagram of binary mixtures. But this
transition can be preempted by the freezing of the large component,
and when this happens the system phase separates into two fcc
solids with a different lattice parameter. This effect, very
clearly shown here for the mixture of PHC (in the limit of infinite
asymmetry), has also been confirmed in simulations of HS interacting
via an effective depletion potential, \cite{noe,dijkstra3} and
very recently also in direct simulations on the true binary
mixture. \cite{dijkstra4} The simulations also show that the
solid-solid transition disappears as the asymmetry of the
two components decreases, but it is anyhow categoric with respect
to the fluid-solid nature of demixing.

A final remark concerns the two-dimensional mixture. We have
made preliminary calculations in this case and have found an
adhesive contribution similar to the three-dimensional one. We
have nor carried out a detailed analysis yet, but the same
collapse is present in this case, thus indicating a behaviour
qualitatively similar to the one shown here, except that we
cannot say anything on the existence of a solid-solid transition.
This results are in perfect qualitative agreement with recent
simulations. \cite{buhot2}

\section*{Acknowledgments}

We like to thank Daan Frenkel, Richard Sear,
and Pedro Tarazona for very useful discussions, and 
Eduardo Jagla for kindly sending us his simulations data. 
JAC's work is part of the research
project PB96--0119 of the Direcci\'on General de Ense\~nanza
Superior (Spain).

\end{multicols}


\begin{thebibliography}{88}
\bibitem[*]{emailyuri} E-mail: \verb+yuri@alum.math.uc3m.es+
\bibitem[\dag]{emailjose} E-mail: \verb+cuesta@math.uc3m.es+
%%%%%%%%%%%%% SECT. INTRODUCTION %%%%%%%%%%%%%%%%%%%%%%%%%%%%%%%%%%%%%%%%%%
\bibitem{cuestaI} J.\ A.\ Cuesta and Y.\ Mart\'\i nez-Rat\'on, \jcp
    {\bf 107}, 6379 (1997).
\bibitem{cuesta1} J.\ A.\ Cuesta and Y.\ Mart\'\i nez--Rat\'on, \prl
    {\bf 78}, 3681 (1997).
\bibitem{zwanzig} R.\ W.\ Zwanzig, \jcp {\bf 24}, 855 (1956).
\bibitem{hoover1} W.\ G.\ Hoover and A.\ G.\ de Rocco, \jcp {\bf 36}, 3141
    (1962).
\bibitem{hoover2} W.\ G.\ Hoover and J.\ C.\ Poirier, \jcp {\bf 38}, 327
    (1963).
\bibitem{vanswol} F.\ van Swol and L.\ V.\ Woodcock, Molec.\ Sim.\ {\bf 1},
    95 (1987).
\bibitem{jagla} E.\ A.\ Jagla, preprint cond-mat/9807032 (1998).
\bibitem{kirkpatrick} T.\ R.\ Kirkpatrick, \jcp {\bf 85}, 3515 (1986).
\bibitem{dijkstra} M.\ Dijkstra and D.\ Frenkel, \prl {\bf 72}, 298 (1994);
     M.\ Dijkstra, D.\ Frenkel and J.-P.\ Hansen, \jcp {\bf 101}, 3179 (1994).
\bibitem{cuesta2} J.\ A.\ Cuesta, \prl {\bf 76}, 3742 (1996).
\bibitem{rowlinson} J.\ S.\ Rowlinson and F.\ Swinton, {\em Liquids and
    Liquid Mixtures} (Butterworths Scientific Publications, London, 1982).
\bibitem{melnyk} T.\ W.\ Melnyk and B.\ L.\ Sawford, Mol.\ Phys.\ {\bf 29},
    891 (1975).
\bibitem{adams} D.\ J.\ Adams and I.\ R.\ McDonald, \jcp {\bf 63}, 1900
    (1975).
\bibitem{ehrenberg} V.\ Ehrenberg, H.\ M.\ Schaink, and C.\ Hoheisel,
    Physica A {\bf 169}, 365 (1990).
\bibitem{carmesin} H.-O.\ Carmesin, H.\ L.\ Frisch, and J.\ K.\ Percus,
    J.\ Stat.\ Phys.\ {\bf 63}, 791 (1991).
\bibitem{lebowitz1} J.\ L.\ Lebowitz, \pra {\bf 133}, 895 (1964).
\bibitem{lebowitz2} J.\ L.\ Lebowitz and J.\ S.\ Rowlinson, \jcp {\bf 41},
    133 (1964).
\bibitem{biben2} T.\ Biben and J.-P.\ Hansen, \prl {\bf 66}, 2215 (1991).
\bibitem{lekkerkerker} H.\ N.\ W.\ Lekkerkerker and A.\ Stroobants, Physica
    A {\bf 195}, 387 (1993).
\bibitem{rosenfeld1} Y.\ Rosenfeld, \prl {\bf 72}, 3831 (1994); J.\
    Phys.\ Chem.\ {\bf 99}, 2857 (1995).
\bibitem{duijneveldt} J.\ S.\ van Duijneveldt, A.\ W.\ Heinen, and H.\ 
    N.\ W.\ Lekkerkerker, Europhys.\ Lett.\ {\bf 21}, 369 (1993).
\bibitem{kaplan} P.\ D.\ Kaplan, J.\ L.\ Rouke, A.\ G.\ Yodh, and D.\ J.\
    Pine, \prl {\bf 72}, 582 (1994); A.\ D.\ Dinsmore, A.\ G.\ Yodh, and
    D.\ J.\ Pine, \pre {\bf 52}, 4045 (1995).
\bibitem{steiner} U.\ Steiner, A.\ Meller, and J.\ Stavans, \prl {\bf 74},
    4750 (1995).
\bibitem{imhof} A.\ Imhof and J.\ K.\ G.\ Dhont, \prl {\bf 75}, 1662 (1995).
\bibitem{poon} W.\ C.\ K.\ Poon and P.\ B.\ Warren, Europhys.\ Lett.\
    {\bf 28}, 513 (1994).
\bibitem{caccamo} C.\ Caccamo and G.\ Pellicane, Physica A {\bf 235},
    149 (1997).
\bibitem{fries} P.\ H.\ Fries and J.-P.\ Hansen, Mol.\ Phys.\ {\bf 48},
    891 (1983).
\bibitem{jackson} G.\ Jackson, J.\ S.\ Rowlinson, and F.\ van Swol, J.\
    Phys.\ Chem.\ {\bf 91}, 4907 (1987).
\bibitem{buhot} A.\ Buhot and W.\ Krauth, \prl {\bf 80}, 3787 (1998).
%\bibitem{dijkstra2} M.\ Dijkstra and R.\ van Roij, \pre {\bf 56}, 5594
%    (1997).
\bibitem{mao} Y.\ Mao, M.\ E.\ Cates, and H.\ N.\ W.\ Lekkerkerker,
    Physica A {\bf 222}, 10 (1995).
\bibitem{biben3} T.\ Biben, P.\ Bladon, and D.\ Frenkel, J.\ Phys.:
    Condens.\ Matter {\bf 8}, 10799 (1996).
\bibitem{noe} N.\ Garc\'\i a-Almarza and E.\ Enciso in {\em Proceedings
    of the VIII Spanish Meeting on Statistical Physics FISES '97}, 
    edited by J.\ A.\ Cuesta and A.\ S\'anchez (Editorial del
    CIEMAT, Madrid, 1998), p.\ 161.
\bibitem{dijkstra3} M.\ Dijkstra, R.\ van Roij, and R.\ Evans,
    \prl {\bf 81}, 2268 (1998).
\bibitem{bolhuis} P.\ Bolhuis and D.\ Frenkel, \prl {\bf 72}, 2211
    (1994); P.\ Bolhuis, M.\ Haagen, and D.\ Frenkel, \pre {\bf 50},
    4880 (1994).
\bibitem{tejero} C.\ F.\ Tejero, A.\ Daanoun, H.\ N.\ W.\ Lekkerkerker,
    and M.\ Baus, \prl {\bf 73}, 752 (1994); \pre {\bf 51}, 558 (1995).
\bibitem{dijkstra4} M.\ Dijkstra, R.\ van Roij, and R.\ Evans,
    preprint (1998).
\bibitem{biben1} T.\ Biben and J.-P.\ Hansen, Europhys.\ Lett.\ {\bf 12},
    347 (1990).
\bibitem{heno} Y.\ Heno and C.\ Regnaut, \jcp {\bf 95}, 9204 (1991).
\bibitem{yuri} Y.\ Mart\'\i nez-Rat\'on and J.\ A.\ Cuesta, \pre
    (in press).
%%%%%%%%%%%%% SEC. ONE-COMPONENT FLUID %%%%%%%%%%%%%%%%%%%%%%%%%%%%%%%%%%%
\bibitem{evans} R.\ Evans in {\em Inhomogeneous Fluids}, edited by D.\
    Henderson (Dekker, New York, 1992).
\bibitem{press} W.\ H.\ Press, S.\ A.\ Teukolsky, W.\ T.\ Vetterling,
    and B.\ P.\ Flannery, {\em Numerical Recipes}, 2nd ed.\ (Cambridge
    University Press, New York, 1992).
\bibitem{nota1} Notice that a FMT functional has no particular difficulties
    to account for vacancies in a crystal lattice, in contrast to what
    happens for most classical functionals. See Ref.\ \onlinecite{evans}
    for a more detailed discussion on this matter.
%%%%%%%%%%%%%% SEC. STABILITY OF THE MIXTURE %%%%%%%%%%%%%%%%%%%%%%%%%%%%%%
\bibitem{callen} H.\ B.\ Callen, {\em Thermodynamics} (Wiley, N.\ Y.,
    1960).
%%%%%%%%%%%%%% SEC. INFINITELY ASYMMETRIC MIXTURE %%%%%%%%%%%%%%%%%%%%%%%%%
\bibitem{stell} G.\ Stell, J.\ Stat.\ Phys.\ {\bf 63}, 1203 (1991).
\bibitem{tejero2} C.\ F.\ Tejero and M.\ Baus, \pre {\bf 48}, 3793 (1993).
\bibitem{frenkel} D.\ Frenkel, private communication.
\bibitem{sear} R.\ P.\ Sear, preprint cond-mat/9806205 (1998).
%%%%%%%%%%%%%% SEC. DISCUSION AND CONCLUSIONS %%%%%%%%%%%%%%%%%%%%%%%%%%%%%
\bibitem{cuestasear} J.\ A.\ Cuesta and R.\ P.\ Sear, Euro.\ Phys.\ J.\
    B (in press).
\bibitem{yurietal} Y.\ Mart\'\i nez-Rat\'on, J.\ A.\ Cuesta, R.\ van
    Roij, and B.\ Mulder in {\em Proceedings of the NATO Advanced
    Study Institute ``New approaches to old and new problems in liquid
    state theory: inhomogeneities and phase separation in simple,
    complex and quantum fluids''}, edited by C.\ Caccamo, J.-P.\ Hansen,
    and G.\ Stell (NATO ASI Series, 1998), in press.
\bibitem{buhot2} A.\ Buhot and W.\ Krauth, preprint cond-mat/9807014 (1998).
\end{thebibliography}
\end{document}